\documentclass{appolb}
\usepackage{epsfig}
\usepackage{amsmath}
\usepackage{amssymb}
\def\runhead{}
\newcommand{\z}[1]{\mathbb{Z}_{#1}}
\newcommand{\R}[1]{\rm{\bf#1}}
\newcommand{\Rb}[1]{\bar{\rm{\bf#1}}}
\newcommand{\Rbb}[1]{\overline{\rm{\bf#1}}}

\newcommand{\I}{{\cal I}}

\newcommand{\iso}{\thickapprox}
\newcommand{\yuk}[1]{Y_{#1}}
\newcommand{\tran}{\Delta}

\begin{document}
\title{Discrete Minimal Flavour Violation%
\thanks{SHEP preprint 09-26  -- Presented at Flavianet workshop on ``Low energy constraints 
on extensions 
of the Standard Model'' from 23-27 July 2009 in Kazimierz, Poland based on \cite{DMFV}.}}
\author{Roman Zwicky$\;^{a}$ \& Thomas Fischbacher$\;^{b}$
\address{$\phantom{x}^a$ School of Physics \& Astronomy,$\phantom{x}^b$  School of Engineering \\
 University of Southampton, Highfield, Southampton SO17 1BJ}}
\maketitle
\begin{abstract}
We investigate the consequences of replacing the continuous flavour  symmetry of minimal flavour 
violation by a discrete group. 
Goldstone bosons,  resulting from the breaking of the continuous flavour symmetry, generically lead  to 
bounds on new flavour structure by many orders of magnitude above the TeV-scale.
The absence of Goldstone bosons for
discrete symmetries constitutes the \emph{primary} motivation of our work. 
The four crystal-like groups  
$\Sigma(168)$, $\Sigma(72 \varphi)$,  $\Sigma(216 \varphi)$ and $\Sigma(360 \varphi)$ provide 
enough protection for a discrete TeV-scale MFV scenario in the case where $\Delta F = 2$ processes
are generated by two subsequent $\Delta F = 1$ transitions.
\end{abstract}
  
\section{Introduction}

In the absence of Yukawa interactions the global flavour symmetry of the standard model (SM) is
$G_F = U(3)^5 = U(3)_Q \times U(3)_{U_R} \times 
U(3)_{D_R} \times U(3)_L \times U(3)_{E_R}$.
It was realized a long time ago \cite{technigim} that  these sort of  flavour symmetries forbid
flavour changing neutral currents (FCNC) at tree-level. The idea behind Minimal Flavour Violation (MFV) \cite{MFV} 
is that the Yukawa matrices are the sole sources, 
\begin{equation}
\label{eq:yuk_break}
G_F = U(3)^5 \stackrel{\yuk{U,D,E}}{\longrightarrow} U(1)_B \times U(1)_L \;,
\end{equation} 
that break this flavour symmetry. 
N.B. the further breaking of this group down to $U(1)_{B-L}$ due to the chiral anomaly \cite{tHooft} 
is not central to this work. For notational simplicity
we shall focus in this work on the quark sector. 
Results can easily be transfered to the lepton sector. 

It is observed that the flavour symmetry is restored 
when the following transformation properties, 
$ \yuk{U} \sim (\R{3}, \Rb{3},\R{1})_{G_q},  \yuk{D} \sim (\R{3},\R{1},\Rb{3})_{G_q}$, are assigned 
to the Yukawa matrices, where 
$G_q = SU(3)_{Q} \times SU(3)_{U_R} \times SU(3)_{D_R}$ is the quark flavour symmetry group\footnote{In this work we take a bit of a cavalier attitude towards U(1) factors. Some remarks on U(1) factrors can be found in \cite{DMFV}}.
An \emph{effective theory} constructed from the SM fields and the Yukawa matrices
is then said to obey  the principle  of \emph{Minimal Flavour Violation} \cite{MFV}, 
if all operators are invariant under $G_q$ \footnote{Sometimes additional assumptions such
as CP invariance \cite{MFV} or no new Lorentz structures  \cite{CMFV} are made.}. The MFV approach is in part motivated by the, so far, pertinent absence of FCNC. Relative bounds 
from $K_0$, $B_d$-oscillations on the Wilson coefficients $C_{\rm SM}/C_{\rm MFV} \simeq  (0.5  {\rm TeV}) ^2 / m_W^2$. The reader is referred to the talks on MFV \cite{christopher} and 
MFV-bounds \cite{federico} for further reference.

Even in the absence of the knowledge of the exact dynamics one delicate 
question might be raised: How is the $G_F$ symmetry broken?
If the symmetry is broken spontaneously, which is what has been proposed so far e.g.
\cite{FJM}, this would then  give rise to $3 \cdot 8 + 2 = 26$  CP-odd massless 
Goldstone bosons, bearing in mind possible U(1) anomalous contributions, in 
the quark sector associated with the breaking of  $U(3)^3 \to U(1)_B$.
Those Goldstone bosons  set 
bounds on new flavour structure many orders of magnitude above the TeV-scale \cite{DMFV}.

In this paper we aim to ameliorate this situation by  replacing $G_F$  by a discrete symmetry\footnote
{Another alternative is to resort to the Higgs mechanism.}.
Spontaneous breaking  of discrete symmetries do \emph{not} lead to Goldstone bosons.
The absence of the latter in this framework  constitutes the \emph{primary} motivation of our 
work. The main remaining issue is then to investigate whether the reduced symmetry provides
enough protection  for a discrete TeV-scale  MFV-scenario.
On the technical side 
the analysis of the effective field theory boils down to the classification of invariants 
of discrete SU(3) subgroups.

\section{Discrete Minimal Flavour Violation}
Replacing  the continuous flavour symmetry with a discrete flavour symmetry requires the following additional information
or assumptions:
\begin{alignat}{2}
\label{eq:discrete}
&a)    \quad  \text{The  group} \quad & & 
D_q  = {\cal D}_{Q_L} \times {\cal D}_{U_R} \times {\cal D}_{D_R} \subset G_q \;, \quad {\cal D} \subset SU(3) \;, \nonumber  \\
&b)  \quad  \text{The representation}  \quad & &  R_3({\cal D}_{Q_L}) \text{ (3D irrep of families )}  \;,  \nonumber  \\
&c)  \quad  \text{Yukawa expansion}  \quad & &  \yuk{U/D} 
\to  \kappa \yuk{U/D} \;, \quad \kappa \in \mathbf{R} \;,
\end{alignat}
where ${\cal D}$ denotes a discrete groups.
The three families transform under a 3D irreducible 
representation (irrep) of ${\cal D}$, which has to be specified since some groups have more than one 
(modulo the complex conjugate).
The Yukawa expansion\footnote{The authors 
\cite{genMFV} distinguish $\kappa \ll 1$ linear MFV versus $\kappa \sim {\cal O}(1)$ non-linear MFV where resummation (non-linear $\sigma$-model techniques) become imperative.}
allows higher dimensional operators to be controlled, which otherwise give rise 
to rather anarchic flavour transitions \cite{DMFV}.

The effective Lagrangian, 
in the absence of the knowledge of the dynamics of the underlying model, is parametrized as,
\begin{equation}
\label{eq:Leff}
{\cal L}^{\rm eff} = \sum_n C^n_{\rm dMFV}   \left(\I_n(\text{Quarks},\text{Yukawas} \right) + h.c)\, , \qquad  C^n_{\rm dMFV}=  \frac{c_n}{  \Lambda^{\text{dim}(\I_n) -4} } \;,
\end{equation}
the sum of all combinations invariant under ${\cal D}_q$ \eqref{eq:discrete}.
The dimension of the operator (invariant) brings in a certain hierarchy in the infinite sum above\footnote{It has to be kept in mind that the association of 
$C^n$ with the scale of new physics is generally obscured by loop factors, mixing angles and renormalization group effects as in any bottom-up effective field theory approach.}. The crucial technical point is that \emph{finding the invariants is equivalent to finding the constant tensors of the symmetry group}. Let us here introduce the following (standard) notation for tensors: 
An index transforming under a $\R{3}$ or  
$\Rb{3}$ representation shall be denoted by 
lower and upper indices respectively
\begin{equation}
\I^{(m,n)} \; \sim \; \I_{\,a_1 .. a_m}^{\,b_1 .. b_n}  \;.
\end{equation}
Non-constant tensors will be denoted by  ${\cal T}^{(m,n)}$.
In principle this tensor classification is not sufficient for our general problem since there are three different group factors \eqref{eq:discrete}. 
It will though prove sufficient here to contract the other indices\footnote{
A refined treatment is only necessary when there
are new $\I^{(2,2)}$ invariants and those groups are not of interest to us anyway.}.
We therefore contract the  ${\cal D}_{U_R}$ index and directly write
 \begin{equation}
 \label{eq:tran}
 (\tran_U) _a^{\phantom{x}r} \equiv ( \yuk{U} \yuk{U}^\dagger)_a^{\phantom{x}r}\;, \qquad     ( \tran 
 \in {\cal T}^{(1,1)}) \;.
\end{equation}
The subscript $U$ shall be dropped when
there is no reason for confusion.
In the reminder we shall use the following notation:
\begin{equation}
\label{eq:para3}
D_L = (d_L , s_L, b_L) \to D_i = (D_1,D_2,D_3) \;, \qquad ( D_i \in {\cal T}^{(1,0)}) \;.
\end{equation}
The following operator classification 
\begin{alignat}{2}
& \I^{(2,2)}_n = (\I_n)^{ab}_{rs}  \, 
\left( \bar D^r  \tran_a^{\phantom{x}s} D_b \right)  \qquad & & \in O^{\Delta F = 1'} \,,
\nonumber \\ 
& \I^{(3,3)}_n =  (\I_n)^{abc}_{rst}  \, 
\left( \bar D^r  \tran_a^{\phantom{x}s} D_b \right) \bar D^t D_c
  \qquad & & \in O^{\Delta F = 1}  \;, \nonumber \\ 
\label{eq:generic2}
&  \I^{(4,4)}_n =  (\I_n)^{abcd}_{rstu}  \, 
\left( \bar D^r  \tran_a^{\phantom{x}s} D_b \right) \, \left( \bar D^t  \tran_c^{\phantom{x}u} D_d \right) \qquad & &  \in O^{\Delta F = 2}   \;,
\end{alignat}
directly connects to the MFV operators \cite{MFV}
\begin{alignat}{1}
\label{eq:DeltaF}
O^{\Delta F = 1'} &=  (\bar D_L \yuk{U} \yuk{U}^\dagger \yuk{D} \, \sigma \!\cdot\! F  D_R)  \;,  
\nonumber \\
O^{\Delta F = 1} &=  (\bar D_L \yuk{U} \yuk{U}^\dagger \gamma_\mu D_L)  \, \bar D_L \gamma^\mu D_L  \;, \nonumber \\ 
O^{\Delta F = 2} &=  (\bar D_L \yuk{U} \yuk{U}^\dagger \gamma_\mu D_L)^2
\;.
\end{alignat}
Before entering into the realm of 
invariants let us briefly discuss the discrete 
SU(3) subgroups.
\subsection{Discrete SU(3) subgroups}
The discrete SU(3) subgroups were classified 
a long time ago \cite{Milleretal} and further analysed as alternatives to SU(3)$_F$ in the context of the eightfold way  \cite{FFK}. Explicit representations and Clebsch-Gordan coefficients were systematically worked out in a  series of papers around 1980 
\cite{BW}, partly motivated as alternatives to $SU(3)_{\rm colour}$ for lattice QCD; and further elaborated very recently 
\cite{simple_discrete} in the context  of family symmetries.

The discrete SU(3) subgroups are of two kinds. 
The analogues of crystal groups, of which we list here the maximal subgroups: 
$\Sigma(168)$,  $\Sigma(360 \varphi)$ and $\Sigma(216 \varphi)$.
The factor $\varphi$ can in general 
either be one or three depending on whether the 
center of SU(3) is divided out or not. For the maximal subgroups it is three.
The second kind are the infinite sequence of groups, sometimes called ``dihedral like'' or ``trihedral'', 
$\Delta(3n^2)   \iso (\z{n} \times \z{n})  \rtimes \z{3}$ and 
$\Delta(6n^2)  \iso (\z{n} \times \z{n})  \rtimes S_3 $ for $n \in {\mathbb Z}$. The symbol 
``$\rtimes$'' denotes a semidirect product. 
The largest irreps of $\Delta(3n^2)$ and 
$\Delta(6n^2)$ are 3-, respectively 6-dimensional; independent of $n$. 

That the catalogue of \cite{FFK} as compared
to \cite{Milleretal} is not complete already surfaced
in the 1980 \cite{BW,response} and
it has recently been reemphasized  \cite{Ludl} that the so-called
$(D)$-groups \cite{Milleretal} have not been discussed systematically in the literature.
In appendix A.2.1 \cite{DMFV} we argue that the 
$(D)$-groups, or more precisely the six-parameter
$D(n,a,b;d,r,s)$ matrix groups, 
are  subgroups of $\Delta(6g^2)$, where
$g$ is the  lowest common multiple of $n$, $d$ and $2$. 
This is sufficient for our consideration.

\subsection{Invariants}
In discussing  invariants we are going to use the  fact, based on the orthogonality theorem, that the 
number of  times the identity appears in a Kronecker product 
$\R{A} \times \R{B} \times  \R{C} \times .. = n_1 \, \R{1} + ..$, denoted by $n_1$, is equal to 
the number of invariants that can be formed out of the irreps $\{ \R{A}, \R{B}, \R{C}, .. \}$\cite{DMFV}. \\[0.2cm]
{\bf New invariants $\I^{(4,4)}$-level -- no $\R{27}$:} The problem of finding all invariants of the $\Delta F = 2$ operator \eqref{eq:DeltaF} is equivalent to finding the invariants of the following Kronecker product: 
$K^{\cal D_{Q_L}} = \left( \Rb{3} \times  \R{3}  \times  \Rb{3} \times \R{3} \right)_s \times \left( \Rb{3} \times
  \R{3}  \times  \Rb{3} \times \R{3} \right)_s$. The symbol  $s$ stands for the symmetric part.
The restriction to the symmetric part can be justified by first considering 
the tensor products of the Yukawas and the quarks separately. The logic that we shall employ is that 
if the Kronecker product of the ${\cal D} \subset SU(3)$ decomposes any different from SU(3) then there are necessarily further invariants. The following Kronecker products will prove sufficient to convey our argument:
\begin{eqnarray}
\label{eq:8}
& &\left( \R{3} \times \Rb{3}\right)_{\rm SU(3)} = 
\R{1} + \R{8} \;, \\
\label{eq:27}
& & (\R{8} \times  \R{8})_{SU(3)} = 
(\R{1} + \R{8} + \R{27})_s  + (\R{8} + \R{10} + \Rbb{10})_a
\end{eqnarray}
The two equations above make it evident that a necessary condition for an identical decomposition 
is that the discrete group contains a 27D irrep.
The trihedral groups  $\Delta(3n^2)$ and $\Delta(6n^2)$ are not in this category since their largest
irreducible representations (irreps) are at most 3-, respectively 6-dimensional. 
Going through the character tables  in \cite{FFK} and the more recent work \cite{Ludl} we realize that there is \emph{no} discrete subgroup  of SU(3) which has a 27D irrep! 
Note, on even more general grounds  that ${\rm dim}(\R{27})^2 = 
729$ almost saturates  the relation between the order of the group and the sum of the dimension of its 
irreps squared \cite{DMFV}  and leaves  $|\Sigma(360 \varphi)|= 1080 > 729 $ as the only hypothetical 
candidate among the crystal-like groups. \\[0.2cm]
{\bf No new invariants $\I^{(2,2)}$-level for four groups:} The good news is though that the  four crystal-like groups
$\Sigma(168)$, $\Sigma(72 \varphi)$,  $\Sigma(216 \varphi)$ and $\Sigma(360 \varphi)$ do have representations that decompose as \eqref{eq:8} and are going to be interesting under the technical assumption of ``family irreducibility'' to be discussed below. Here we shall give an overview of the number of complex conjugate 3D irreps and 
the number of invariants (under certain symmetrizations):
\begin{table}[h]
\begin{center}
\begin{tabular}{l | r |  c | c | c c | c c c |}
group &  order & $(\R{3},\Rb{3})$ & $\I^{(2,2)}$ & $\I^{(3,3)}$ & $\I^{(3,3)}_{2,1}$  & $\I^{(4,4)}$ & $\I^{(4,4)}_{3,1}$ & $\I^{(4,4)}_{2,2}$  \\[0.1cm]
\hline
& & & & & & & & \\[-0.3cm]
SU(3)                                     & $\infty$ \quad   & 1& 2  & 6   & 5 & 23 & 15 & 14 \\
$\Sigma(360\varphi)$            & 1080  \quad  & 2 & 2  & 6  &  5 &  28 & 18 & 17 \\
$\Sigma(216\varphi)$            & 648  \quad &3 &2  & 7  &  6 &  40 & 27 & 23 \\
$\Sigma(168)$                       & 168  \quad &1 & 2  & 7   & 6 & 44 & 29 & 25 \\
$\Sigma(72\varphi)$              & 216  \quad &4 & 2  & 11 & 8 &  92 & 55 & 43 \\
\end{tabular}
\end{center}
\caption{\sl \small  Number of invariants for tensors of type $\I^{(n,n)}$, whose definition 
can be inferred from Eq.~\eqref{eq:generic2}. 
The subscripts $x,y$ indicate symmetrizations of $x$ and $y$ pairs of 
$\R{3},\Rb{3}$ indices. $\I^{(3,3)}_{2,1}$ and $\I^{(4,4)}_{2,2}$ correspond
to the (symmetric) contractions of
$\Delta F = 1$ and $\Delta F =2$ in \eqref{eq:generic2}.} 
\label{tab:candidates}
\end{table}
Note that we implicitly stated that there are
groups where $\Delta F =2$ operators are possible with $\I^{(2,2)}$ solely; without any Yukawa transitions.
The group $\Sigma(60) \iso A_5$ is an example which is discussed in appendix A of reference \cite{DMFV}.

\begin{figure}[h]
 \centerline{\includegraphics[width=4.0in]{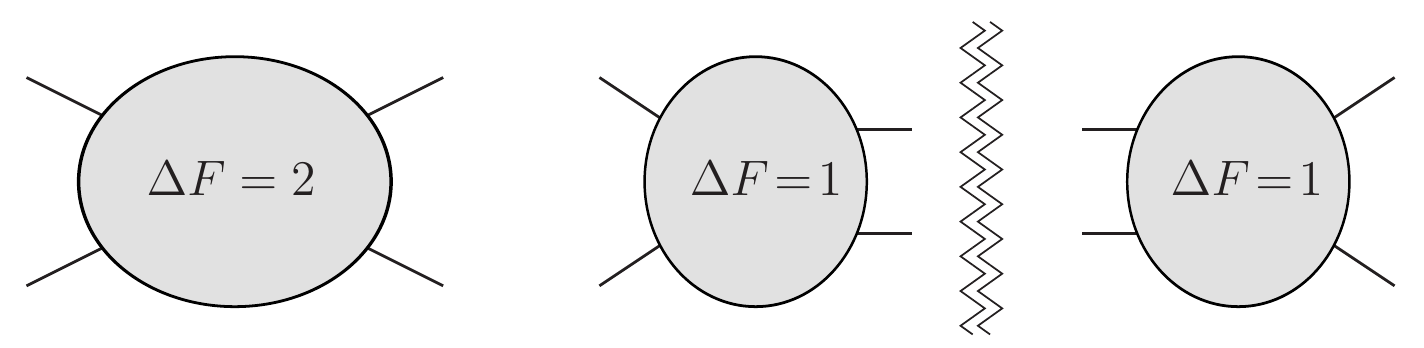}
 }
 \caption{\small (Left) ``family irreducible''
 (Right)  ``family reducible''}
\label{fig:DeltaF12}
\end{figure}

It is argued, with concrete examples, that new invariants necessarily upset the flavour hierarchy of MFV \cite{DMFV}. 
In particular the breaking of the continuous flavour group down to a discrete group implies that 
the mass-flavour basis transformation become observable and deprive the approach of its predictivity.
To conclude, that the necessity of new invariants at the $\I^{(4,4)}$-level  and its connection with 
the well-tested $\Delta F = 2$ transitions, imply that no discrete flavour group is suitable
appears too hasty.
The generation mechanism of  $\Delta F = 2$ operators has to be reflected upon.
We distinguish the two cases where the $\Delta F =2$ process 
is generated via two
subsequent $\Delta F = 1$ parts and where
this is not the case. We  shall call the former  ``family irreducible'' 
and the latter ``family reducible'',  c.f. Fig.~\ref{fig:DeltaF12}.  
The SM and the R-parity conserving MSSM are examples of the ``family reducible''-type whereas
 the R-parity violating MSSM is of the  ``family irreducible''-type.  
 The composite technicolor model \cite{gc}, in the absence of the knowledge of its
 non-perturbative dynamics, have to be counted into the latter class as well.
 ``Family reducability'' essentially implies
\begin{equation}
(\Delta F= 2) \iso (\Delta F =1) \times (\Delta F =1) \; \Rightarrow \; \I^{(4,4)} \to \I^{(2,2)}\I^{(2,2)} \;,
\end{equation}
factorization of the invariant and proves to be a \emph{sufficient} condition for a discrete TeV-scale scenario 
for the four crystal-like groups listed in Tab.~\ref{tab:candidates}. To this end we would like to discuss two further
points:
\begin{itemize}
\item$\Sigma(360\varphi)$ model-independent:
The most suitable candidate, for a TeV-scale dMFV scenario is $\Sigma(360\varphi)$,
since the first new invariants appear only at the $\I^{(4,4)}$-level c.f. Tab~\ref{tab:candidates}. 
Yet: \emph{How small does $\kappa$ need to be in order for $C_{\rm dMFV}$ \eqref{eq:Leff} to satisfy the same
kind of experimental bounds as for $C_{\rm MFV}$  found in reference \cite{MFV}?}
The discussion in section  5.1 \cite{DMFV}  suggests that 
$s \to d$ could be induced at first order in 
$\lambda$ for new invariants, as compared to  order
$|V_{ts} V_{td}^*| \sim \lambda^5$ in MFV.
A  $\Delta S = 2$ transition could therefore 
be ${\cal O}(\lambda^6)$ as compared to 
${\cal O}(\lambda^{10})$ in MFV.
According to our reflection above we have to balence:
$ {\rm MFV} : {\rm DMFV}  = \lambda^{10} : \lambda^{6} \kappa^4 \Rightarrow 
\kappa_{\Sigma(360\varphi)} \simeq \lambda \simeq 0.2$.
\item It has to be kept in mind that even when a model is in the ``family irreducible''-class,
the fact that the
vertices are ${\cal D}_q$-invariant prevents the generation of 
non factorizable $\I^{(4,4)}$-invariants. We argue that this is indeed the case for the R-parity violating 
MSSM (at least to leading order) \cite{DMFV}.
\end{itemize}
Our aim, in this work, was to point out general issues of implementing 
MFV via a discrete group. We would hope that this work would be of some help
for further investigations towards more specific models. 
It also has to be said that although MFV has very attractive features, e.g. a sufficiently stable
proton in the R-parity violating MSSM \cite{proton-decay}, so far no explicit model
for MFV has appeared in the literature.
Moreover the formulation \eqref{eq:discrete}  could be 
refined by constructing a discrete subgroup of 
$G_q$  which does not factor into 
direct products of SU(3) subgroups \cite{DMFV}. 
One might wonder what the consequences are,
of such a non-trivial embedding, for the invariants.

{\bf Acknowledgements:} 
RZ would like to thank many colleagues for inspiring discussions and remarks 
and at last I would like to thank the organizers if the Kazimierz workshop for the organization and a very pleasant 
atmosphere. RZ acknowledges the support of an advanced STFC fellowship.

\end{document}